\documentclass[review]{elsarticle}

\usepackage{hyperref}
\usepackage{amsmath,amssymb,amsfonts}
\usepackage{bm}
\usepackage{makecell}

\journal{Journal of \LaTeX\ Templates}

\begin{document}

\begin{frontmatter}

\title{PMR-Net: Parallel Multi-Resolution Encoder-Decoder Network Framework for Medical Image Segmentation}
\tnotetext[mytitlenote]{Fully documented templates are available in the elsarticle package on \href{http://www.ctan.org/tex-archive/macros/latex/contrib/elsarticle}{CTAN}.}

\author[myfirstaddress,mysecondaryaddress]{Xiaogang Du}

\author[myfirstaddress,mysecondaryaddress]{Dongxin Gu}

\author[myfirstaddress,mysecondaryaddress]{Tao Lei\corref{mycorrespondingauthor}}
\cortext[mycorrespondingauthor]{Corresponding author}
\ead{leitao@sust.edu.cn}

\author[myfirstaddress,mysecondaryaddress]{Yipeng Jiao}

\author[myfirstaddress,mysecondaryaddress]{Yibin Zou}


\address[myfirstaddress]{Shaanxi Joint Laboratory of Artificial Intelligence, Shaanxi University of Science and Technology, Xi’an, China}
\address[mysecondaryaddress]{School of Electronic Information and Artificial Intelligence, Shaanxi University of Science and Technology, Xi’an, China}

\begin{abstract}
In recent years, popular encoder-decoder networks consider expanding the receptive field on deep feature maps and integrating multi-scale context semantic features, so as to extract global features suitable for objects with different sizes. However, with the deepening of networks, a large number of spatial fine-grained features are discarded, which makes the networks unable to locate the objects accurately. In addition, the traditional decoder continuously performs upsampling through interpolation, which makes the global context feature lost and reduces the segmentation accuracy at the object edge. To address the above problems, we propose a novel parallel multi-resolution encoder-decoder network, namely PMR-Net for short. First, we design a parallel multi-resolution encoder and a multi-resolution context encoder. The parallel multi-resolution encoder can extract and fuse multi-scale fine-grained local features in parallel for input images with different resolutions. The multi-resolution context encoder fuses the global context semantic features of different receptive fields from different encoder branches to maintain effectively the integrity of global information. Secondly, we design a parallel multi-resolution decoder symmetrical to the structure of parallel multi-resolution encoder. The decoder can continuously supplement the global context features of low-resolution branches to the feature maps of high-resolution branches, and effectively solve the problem of global context feature loss caused by upsampling operation in the decoding process. Extensive experiment results demonstrate that our proposed PMR-Net can achieve more accurate segmentation results than state-of-the-art methods on five public available datasets. Moreover, PMR-Net is also a flexible network framework, which can meet the requirements of different scenarios by adjusting the number of network layers and the number of parallel encoder-decoder branches.
\end{abstract}

\begin{keyword}
\texttt{Medical image segmentation, Deep learning, Convolutional neural network, Multi-scale features, Context features}
\end{keyword}

\end{frontmatter}


\section{Introduction}
Medical image segmentation is a procedure of extracting the region of interest from medical images according to some characteristics. It plays a very important role in several fields of medical image analysis, such as volumetric analysis and measurement, longitudinal analysis, and cortical surface analysis. In practical applications, medical image segmentation is also important for some clinical procedures such as radiotherapy, image-guided surgery, and pathological diagnosis et al [1, 2].

In earlier years, medical image segmentation mainly relied on model-driven based image segmentation methods such as region growing [3], active contour models [4], shape statistical models [5], fuzzy clustering [6], etc. Generally, these model-driven based image segmentation methods do not need a lot of data to train a model, and these methods have good theoretical basis and interpretability. However, the segmentation accuracy and robustness of these methods still requires to be improved for objects in complex scenes, such as tumors with intensity inhomogeneity or blur edges.

Recently, with the rapid development of deep learning techniques, convolutional neural networks (CNNs) have attracted enormous attention in medical image segmentation [7]. Due to the good capacity of the feature representations, CNN based methods have been widely applied in various medical imaging modalities such as X-ray, computed tomography (CT), magnetic resonance imaging (MRI), endoscopy, dermoscopy, and electron microscopy. Currently, popular segmentation networks usually depend on the encoder-decoder structure [8-9]. The success of encoder-decoder based networks is largely attributed to the use of skip connections, which allows the propagation of feature maps from the encoder to the decoder [10]. However, existing encoder-decoder based networks cannot capture and maintain simultaneously the multi-scale features suitable for objects with different sizes, thus it is very difficult to segment objects with different sizes accurately and completely. Moreover, the local features are progressively restored by the upsampling operation in the decoding stage, but the context features containing global information are lost, resulting in the inaccurate segmentation results.

To address the aforementioned issues, we propose a parallel multi-resolution encoder-decoder network, called PMR-Net for short. The main contributions of this work are summarized as follows.
\begin{itemize}
\item[1)]
We design a parallel multi-resolution encoder, which effectively improves the multi-scale feature representations by extracting and fusing the local and global features from the multi-resolution input images. Additionally, a multi-resolution context encoder is designed to extract effectively multi-scale context features from multi-resolution input images.
\end{itemize}

\begin{itemize}
\item[2)]
We design a parallel multi-resolution decoder symmetrical to the structure of parallel multi-resolution encoder. The parallel multi-resolution decoder can implement the fusion of global and local features and effectively alleviate the loss of global features due to upsampling in the decoding stage.
\end{itemize}

\begin{itemize}
\item[3)]
Extensive experiments are carried out on five public available datasets and the experimental results demonstrate that the proposed PMR-Net is superior to the state-of-the-art methods. More importantly, PMR-Net can flexibly adjust the number of network layers and parallel encoder-decoder branches, and can be applied to medical image segmentation with different scenes.
\end{itemize}

The rest of this paper is organized as follows. Section 2 describes the related work of medical image segmentation. Section 3 puts forward PMR-Net and introduces the implementation details of PMR-Net. Section 4 demonstrates the segmentation results of PMR-Net on five public datasets. Section 5 discusses the flexibility in structural design of the PMR-Net. Section 6 concludes the paper.

\section{Related work}
In the past decade, many algorithms have been developed and continuously improved for medical image segmentation. Ronneberger et al [11] proposed the U-Net, which plays an important role in promoting the development of medical image segmentation. Moreover, many popular methods such as V-Net [12], UNet++ [13] and Graph U-Nets [14] are implemented based on U-Net. These methods mainly include three modules: encoder, context encoder and decoder. In this section, we review the related work from three folds: encoders, context encoders and decoders.

\subsection{Encoders}
An encoder can extract many different features to solve the task of medical image segmentation. The shallow stage of an encoder can extract fine-grained features, such as texture features. With the continuous pooling, the deep stage of the encoder can extract rich global context features. To strengthen the feature extraction ability of encoder, scholars continue to increase the number of network layers. However, with the increase of network layers, the gradient is easy to disappear or explode in the process of network training, which makes the network difficult to converge. To address this issue, both the residual network [15] and the dense network [16] propose shortcut connection to alleviate network degradation, and researchers employ shortcut connection to medical image segmentation network [17-19]. The use of shortcut connection can improve the network performance, but fine-grained features are still lost due to pooling or convolution with large strides.

To solve this problem, scholars introduced convolution kernel with different receptive fields instead of vanilla convolution kernel with fixed receptive fields to extract features. For instance, Wang et al. [20] proposed global aggregation blocks in non-local U-Nets to solve the problem of low efficiency caused by the limitation of the receptive field of convolution. Chen et al. [21] introduced Transformers with global self-attention mechanisms as alternative architectures to model long-range dependency. Additionally, inspired by the Inception structure [22-23], Su et al. [24] utilized convolution kernels with different receptive fields to extract multi-scale features in parallel, which improves the feature representation ability but increases the parameters of the network.

To extract multi-scale features without significantly increasing the parameters of network, scholars used many small convolution kernels instead of single large convolution kernel. For example, Gao et al. [25] proposed Res2Net module based on group convolution strategy. This module employs two $3\times3$ convolutions instead of one $5\times5$ convolution and three $3\times3$ convolutions instead of one $7\times7$ convolution, respectively, which can improve the multi-scale feature representations of the model more effectively. Ibtehaz et al. [26] put forward MultiResUNet, which also uses many small convolutions instead of single large convolution, and changes the parallel multi-branch structure to single branch structure to extract multi-scale features. Xie et al. [27] and Peng et al. [28] also realized the idea of reducing the model parameters while extracting multi-scale features. However, the multi-scale features extracted by these methods can not be effectively retained after continuous pooling operation. Therefore, they still can not accurately segment objects with different sizes.

To address this problem, Liu et al. [6] used multi-scale input images to fuse the multi-scale information and proposed MDAN-UNet. In this model, the original images of different scales are added to each layer of the encoder to supplement the multi-scale features. Salehi et al. [29] adopted the multi-scale input image strategy to the 3D medical image segmentation task and proposed Auto-Net. The model cuts input images into blocks according to different window sizes and extracts features in parallel, which can utilize the local and global features of images leading to the improvement of the segmentation accuracy.

\subsection{Context Encoders}
An context encoder can extract high-level global context features. To make better use of the context information to accurately segment objects with different sizes, scholars have carried out a lot of researches and improvements on the context encoder. Zhao et al. [30] adopted a pyramid pooling module, which can extract global context features through context information aggregation based on different regions, and combines them with local features to jointly predict more reliable segmentation results. Chen et al. [31, 32] successively proposed atrous convolution and atrous spatial pyramid pooling module, which utilizes atrous convolution with different dilated rates to capture multi-scale context features in parallel. Inspired by the Inception-ResNet, Gu et al. [33] design dense atrous convolution module and residual multi-scale pooling module and apply them to medical image segmentation. Compared with the traditional convolution, atrous convolution can expand the receptive field without increasing the number of parameters, and then extract high-level contextual semantic features to segment objects with different sizes more effectively.

However, the above methods still have two problems. Firstly, these methods pay the same attention to important and unimportant features, resulting in adding a lot of useless information to the extracted features, and even reducing the model performance. Secondly, it is easy to degrade the model performance due to using atrous convolution with large dilated rate. To solve the first problem, Fu et al. [34] employed the attention mechanism [35] to adaptively extract global context semantic features. He et al. [36] proposed global guided local affinity based on the attention mechanism. This method also takes into account the global features of images while assigning adaptive weights to the local features of pixels. Aiming at the second problem, Yang et al. [37] densely connected the atrous convolution and put forward DenseASPP, which can not only extract and fuse a larger scale range of multi-scale features, but also alleviate the performance degradation of the model caused by atrous convolution with large dilated rate.

\subsection{Decoders}
In classical medical image segmentation models such as UNet [11], SegNet [38] and DeconvNet [39], the high-level semantic features extracted by the encoder need to be upsampled layer by layer by the decoder, and then the segmentation probability maps with the same size as the original image are output. However, the decoder used in these classical models usually outputs coarser segmentation results, and the positioning effect at the object edge is poor. In order to improve the positioning accuracy of the model at the object edge, scholars introduced a post-processing method at the end of the deep neural network model. For example, Chen et al. [31] employed the conditional random field after DeepLab network to further process the coarse segmentation results, which can effectively improve the positioning performance at the object edge. Although introduction of post-processing improves the final segmentation results, it is not an end-to-end design scheme.

With the development of deep learning, network cascading [40-43] and stacking [44-46] provide two effective strategies for end-to-end design. Network cascading is to decompose tasks into different sub-tasks, and then use cascaded sub-networks to deal with sub-tasks with different difficulties. Network stacking repeatedly processes the coarse segmentation results from bottom to top and from top to bottom, so as to improve the segmentation results from coarse to fine. Network cascading and stacking can effectively improve the segmentation accuracy, but they will lead to a large number of model parameters. To extract fine-grained features without increasing the number of model parameters, scholars adopted the preservation strategy of high-resolution feature maps to improve the feature recovery ability of the model [47-49]. For example, Wang et al. [47] retained the high-resolution feature maps in the encoding and decoding process of HR-Net, and extracted and fused multi-scale features on parallel branches. This method improves the segmentation performance of HR-Net without significantly increasing the model parameters.

\section{Method}
In this section, we first describe the overall architecture of the proposed PMR-Net. Then, we design three important components of PMR-Net: parallel multi-resolution encoder, multi-resolution context encoder and parallel multi-resolution decoder. Finally, we introduce the loss function used in PMR-Net.

\subsection{Network architecture}
To segment accurately objects with different sizes, we introduce the strategy of parallel feature extraction and propose PMR-Net based on UNet++. PMR-Net is a parallel multi-resolution encoder-decoder network mainly including three important components: the parallel multi-resolution encoder, the multi-resolution context encoder, and the parallel multi-resolution decoder. The overall architecture of PMR-Net is illustrated in Figure 1.
\begin{figure*}[htbp]
	\centering
	\includegraphics[width=1.2\linewidth]{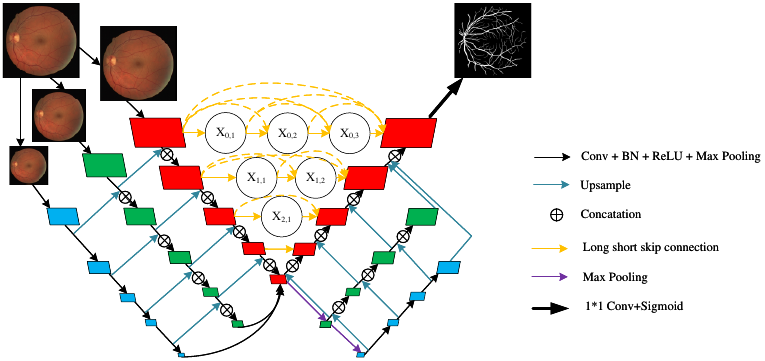}
	\caption{The network structure of PMR-Net.}
\label{fig1}
\end{figure*}

Firstly, we design a parallel multi-resolution encoder to extract multi-scale features. The parallel multi-resolution encoder has two advantages. The first is that the encoder does not use any pre-trained deep network as the backbone, which widens the network instead of deepening the network like ResNet and alleviates the problem of vanishing/exploding gradients for stochastic gradient descent(SGD) with back propagation. The second is that only $3\times3$ convolutions are employed in the parallel branches of the encoder to extract multi-scale features, which can effectively reduce the parameters of PMR-Net. Image $x_1$ and $x_2$ are obtained after downsampling the original input image $x_0$ once and twice, respectively. Image $x_0$,  $x_1$, and $x_2$ are input as three parallel branches of the encoder, respectively. Secondly, three branches of the encoder extract the feature maps $f(x_0)$,  $f(x_1)$, and $f(x_2)$, which are encoded and decoded in the multi-resolution context encoder to fuse the global and local features and obtain richer high-level semantic features. The fused feature maps are pooled twice to obtain the feature maps with three different resolutions. Finally, to recover the objects more accurately, we design a parallel multi-resolution decoder symmetrical to the structure of parallel multi-resolution encoder. In the decoding stage, the lost global features due to upsampling are continuously supplemented to the restored feature maps with the original size, and the feature maps of low-resolution branch are supplemented to the feature maps of high-resolution branch. Thus, it can effectively help the accurate restoration of local features by supplementing global features.

\subsection{Parallel multi-resolution encoder}
Image $x_1$ and $x_2$ are obtained by downsampling the original image once and twice, respectively. Image $x_0$, $x_1$, and $x_2$ are used as the input of parallel multi-resolution encoder to extract features layer by layer in parallel. The size of the input image of each branch is different, so the semantic features extracted by each branch are different. The branch of high-resolution image focuses on extracting local features, and the branch of low-resolution image focuses on extracting global features. Compared with other encoder using convolutions with different receptive fields, parallel multi-resolution encoder can effectively extract features with different scales and effectively reduce the feature redundancy. In the encoding stage, the global feature maps generated by the low-resolution image are fused with the local feature maps generated by the high-resolution image. The structure of the parallel multi-resolution encoder is shown in Figure 2.
\begin{figure}[!t]
\centerline{\includegraphics[width=0.85\columnwidth]{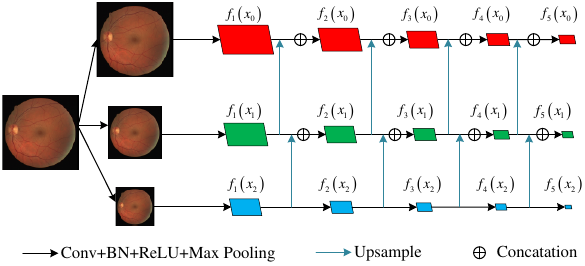}}
\caption{Parallel multi-resolution encoder.}
\label{fig2}
\end{figure}

In Figure 2, the three branches of the parallel multi-resolution encoder respectively extract the feature maps $f_{i}(x_0)$,  $f_{i}(x_1)$, and $f_{i}(x_2)$, where $i$ indicates the layer number of network and \{$i \in N | 1 \leq i \leq 5$\}. Generally, the high-resolution branch focuses on extracting and retaining local features, and the low-resolution branch focuses on extracting and retaining the global features through expanding the receptive field. Meanwhile, Three feature fusion steps are carried out progressively in the encoding stage. First, The feature map $f_{i+1}(x_2)$ is obtained by performing convolution operation on the feature map $f_{i}(x_2)$. Second, $f_{i+1}(x_1)$ is obtained by the fusion of $f_{i}(x_2)$ and $f_{i}(x_1)$. Third, $f_{i+1}(x_0)$ is obtained by the fusion of $f_{i}(x_1)$ and $f_{i}(x_0)$. Here $f_{i}(x_2)$ is upsampled by bilinear interpolation to be the same size as $f_{i}(x_1)$, and $f_{i}(x_1)$ is also upsampled to be the same size as $f_{i}(x_0)$. Therefore, the feature maps extracted from the three branches are progressively fused in each layer of the encoder. With the fusion of feature maps with different semantic information, parallel multi-resolution encoder can extract much richer multi-scale features. Feature fusion in parallel multi-resolution encoder is defined as:
\begin{equation}\label{1}
  f_{i+1}(x_j)=\left\{ \begin{aligned}
PRBC(f_{i}(x_{j}) \oplus U(f_{i}(x_{j+1})))  & & j=0 \\
PRBC(f_{i}(x_{j}) \oplus U(f_{i}(x_{j+1})))  & & j=1 \\
PRBC(f_{i}(x_{j}))  & & j=2\\
\end{aligned} \right.,
\end{equation}
where $i$ refers to the number of layer, and \{$i \in N | 1 \leq i \leq 4$\}. $U$ stands for the upsampling operation using bilinear interpolation. $\oplus$ represents the concatenation operation on the channel dimension of feature maps. $PRBC$ denotes the composite operation of convolution, BN, ReLU, and max pooling in sequence.

\subsection{Multi-resolution context encoder}
To integrate fully the multi-scale context features extracted from the three branches of the parallel multi-resolution encoder, we design a multi-resolution context encoder, and take the multi-scale context features extracted by the parallel multi-resolution encoder as the input of the multi-resolution context encoder. The designed multi-resolution context encoder is illustrated in Figure 3.
\begin{figure}[!t]
\centerline{\includegraphics[width=0.6\columnwidth]{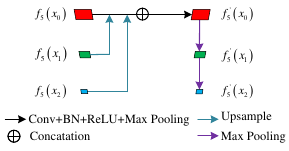}}
\caption{Multi-resolution context encoder.}
\label{fig3}
\end{figure}

In Figure 3, Firstly, the feature maps $f_{5}(x_2)$ and $f_{5}(x_1)$ are upsampled to make them the same size as the feature maps $f_{5}(x_0)$. Then, the upsampling results of $f_{5}(x_2)$ and $f_{5}(x_1)$ are fused with $f_{5}(x_0)$, and the fusion results are used as the input of multi-resolution context encoder. The feature encoding process of multi-resolution context encoder is described as:
\begin{equation}\label{2}
f_{5}^{'}(x_0)=RBC(U(f_{5}(x_2)) \oplus U(f_{5}(x_1)) \oplus(f_{5}(x_0))),
\end{equation}
where $U$ represents the upsampling operation using bilinear interpolation. $\oplus$ represents the concatenation operation on the channel dimension of feature maps. $RBC$ represents the composite operation of convolution, BN, and ReLU in sequence.

In Figure 3, multi-scale context features are integrated in the encoding stage of the multi-resolution context encoder. Then, the multi-scale context features are decoded to generate three kinds of different scale semantic features, which are transmitted to the parallel multi-resolution decoder. The feature decoding of multi-resolution context encoder is represented as:
\begin{equation}\label{3}
\left\{ \begin{aligned}
f_{5}^{'}(x_1)=Pool(f_{5}^{'}(x_0)) \\
f_{5}^{'}(x_2)=Pool(f_{5}^{'}(x_1)) \\
\end{aligned} \right..
\end{equation}

According to (3), max pooling is carried out for the feature maps $f_{5}^{'}(x_0)$ to obtain $f_{5}^{'}(x_1)$ and $f_{5}^{'}(x_2)$. The fused multi-scale context features are decoded into new feature maps with different sizes. After the context features of each branch are encoded and decoded, the local feature maps $f_{5}^{'}(x_0)$ extracted from high-resolution branches are fused with global features, and the global feature maps $f_{5}^{'}(x_1)$ and $f_{5}^{'}(x_2)$ extracted by low-resolution branches are fused with local features. Three feature maps $f_{5}^{'}(x_0)$, $f_{5}^{'}(x_1)$, and $f_{5}^{'}(x_2)$ are transmitted to the parallel multi-resolution decoder for subsequent decoding processing.

Compared with the dense atrous convolution (DAC) module in CE-Net, the proposed multi-resolution context encoder only uses $3\times3$ convolution instead of $1\times1$ convolution and $3\times3$ atrous convolutions with different dilated rates in the DAC module, and fully integrates the extracted semantic features with different scales. It effectively alleviates the loss of local information caused by using atrous convolution to expand the receptive field.

\subsection{Parallel multi-resolution decoder}
In the upsampling process of classical decoder, the global context features are gradually lost when the detailed features are gradually restored. To solve this problem, we designed a parallel multi-resolution decoder symmetrical to the parallel multi-resolution encoder structure, as depicted in Figure 4.

In Figure 4, parallel multi-resolution decoder mainly includes three key steps. First, $f_{i}^{'}(x_2)$ is upsampled by bilinear interpolation in the upsampling process of parallel multi-resolution decoder, and the $3\times3$ convolution kernel is used to fuse the upsampling results with the feature maps $f_{i}^{'}(x_1)$. Secondly, the feature maps $f_{i}^{'}(x_1)$ is upsampled by bilinear interpolation, and the upsampling results are fused with the feature maps $f_{i}^{'}(x_0)$. Feature fusion is carried out layer by layer in the whole decoding stage. Through the above steps, the global context features extracted from low-resolution branches are fused into the feature maps extracted from high-resolution branches. Finally, to improve the segmentation accuracy of PMR-Net, the long and short skip connections are introduced to fuse feature maps from the parallel multi-resolution encoder, and the fused feature maps $f_{i}^{*}(x_0)$ are added to the parallel multi-resolution decoder. In the parallel multi-resolution decoder, the feature maps $f_{i}^{'}(x_j)$ are computed in (4):
\begin{figure}[!t]
\centerline{\includegraphics[width=0.85\columnwidth]{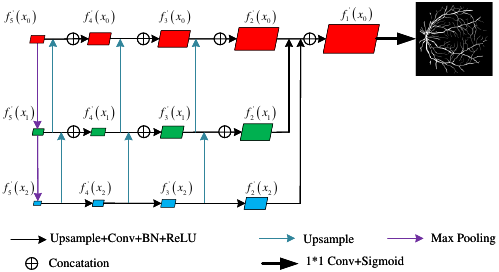}}
\caption{Parallel multi-resolution decoder.}
\label{fig4}
\end{figure}

{\footnotesize
\begin{equation}\label{4}
  f_{i}^{'}(x_j)=\left\{ \begin{aligned}
URBC(f_{i+1}^{'}(x_{j}) \oplus U(f_{i+1}^{'}(x_{j+1}))U(f_{i}^{*}(x_{j})) & & j=0 \\
URBC(f_{i+1}(x_{j}) \oplus U(f_{i+1}(x_{j+1})))  & & j=1 \\
URBC(f_{i+1^{'}}(x_{j}))  & & j=2\\
\end{aligned} \right.,
\end{equation}
}
where $i$ refers to the number of the layer, and \{$i \in N | 1 \leq i \leq 4$\}. $U$ stands for the upsampling operation using bilinear interpolation. $\oplus$ represents the concatenation operation on the channel dimension of feature maps. $URBC$ denotes the composite operation of convolution, BN, ReLU, and upsampling in sequence. Finally, parallel multi-resolution decoder reduces the number of channels by $1\times1$ convolution, generates the segmentation probability maps by Sigmoid function, and then gets the final segmentation results by threshold processing.

\subsection{Loss function}
Image segmentation needs to predict whether each pixel belongs to the foreground or background. Binary cross entropy can effectively compute the loss between the predicted value and the ground truth. It is often used as the loss function of model for two classification tasks. However, the object only accounts for a small part of the whole image in the small object segmentation task. Meanwhile, the loss of calculating image background accounts for more proportion in binary cross entropy, and the network parameters learned in the training process tend to pay more attention to the segmentation performance of image background. Therefore, binary cross entropy is not optimal as the loss function of models for the small objects segmentation. Fortunately, Dice coefficient is a measure of the overlap of foreground areas, which is suitable for evaluating the model performance for small objects segmentation. Consequently, in order to better converge the loss function and improve the segmentation accuracy of small object segmentation tasks, binary cross entropy and dice loss are jointly used as the loss function to conduct the training of PMR-Net in this paper. The loss function of PMR-Net is defined as:
\begin{equation}\label{5}
\mathcal{L}oss= \frac{1}{2} \times\mathcal{L}oss_{BCE} + \mathcal{L}oss_{Dice},
\end{equation}
where $\mathcal{L}oss_{BCE}$ represents the loss of binary cross entropy, which is computed in (6):
\begin{equation}\label{6}
\mathcal{L}oss_{BCE}= -\frac{1}{N}\sum_{i=1}^N (y_{i} \times \log_{2}\widehat{y_{i}} + (1-y_{i}) \times \log_{2}(1-\widehat{y_{i}})),
\end{equation}
where $N$ denotes the total number of pixels of the image. $y_{i} \in \{0,1\}$ and \{$\widehat{y_{i}} \in R | 0 \leq \widehat{y_{i}} \leq 1\}$ represent the ground truth and predicated value, respectively. The Dice loss is calculated in (7):
\begin{equation}\label{7}
\mathcal{L}oss_{Dice}= 1-\frac{2 \times \sum_{i=1}^N y_{i} \times \widehat{y_{i}}}{\sum_{i=1}^N y_{i}^{2}+\sum_{i=1}^N \widehat{y_{i}^{2}} + smooth}.
\end{equation}

In (7), $smooth$ is set to $10^{-5}$, which is introduced to prevent the denominator from being zero.

\section{Experiments}
In this section, we first describe the experiment setup, including datasets, evaluation criteria and implementation details. Then, we demonstrate the performance of PMR-Net and compare it with previous state-of-the-art methods. Finally, we conduct a series of ablation studies to highlight the impact of each component of PMR-Net on the performance.

\subsection{Experiment Setup}
\paragraph{Datasets}To effectively evaluate the performance of PMR-Net, we carry out comprehensive experiments on five commonly used datasets including lung [50], retinal vessel (DRIVE) [51], retinal vessel (STARE) [52], skin lesions (SL) [53-54], and cell  nucleus [55] datasets, respectively. These datasets are listed in Table $1$.

\begin{table}[h]\centering
\caption{The image segmentation datasets used in experiments.}
\label{table}
\setlength{\tabcolsep}{3pt}
\begin{tabular}{ccccc}
\hline
Datasets&Images&Image size&Modality&Provider\\ 
\hline
Lung [50]&267&512$\times $512&X-ray&LUNA challenge\\
Retinal vessel [51]&40&584$\times $565&OCT&DRIVE\\
Retinal vessel [52]&20&700$\times $605&OCT&STARE\\
Skin lesions [53-54]&2694&2166$\times $3188&Dermoscopy&ISIC 2018\\
Cell nuclei [55]&670&256$\times $256&EM&DSB 2018\\
\hline
\end{tabular}
\label{tab1}
\end{table}

(1)\ Lung dataset [50]. From the Lung Nodule Analysis (LUNA) Challenge, which involves processing and trying to find nodular areas in lung CT. In order to find the diseased regions well in these images, the lungs are first needed to be segmented. The image size is $512\times512$ and the number of samples is $267$.

(2)\ Retinal vessels (DRIVE) dataset [51]. This dataset consists of $40$ color fundus images. The image size is $584\times565$. $40$ images were randomly divided equally, $20$ for training and $20$ for testing.

(3)\ Retinal vessels (STARE) dataset [52]. The dataset consists of $20$ images of retinopathy with a macula with an image size of $700\times605$. They were manually annotated by two ophthalmologists, Adam Hoover and Valentina Kouznetsova. We chose Adam Hoover's manual annotations as training labels. Compared with the DRIVE dataset, this dataset is more difficult to be segmented and has more clinical value.

(4)\ Skin lesions dataset [53-54]. The International Skin Imaging Collaboration (ISIC) released a lesional skin dataset including $2694$ images. The average size of images is $2166\times3188$. $2594$ images are chosen for training, and $100$ images are chosen for evaluation.

(5)\ Cell nuclei dataset [55]. This dataset, provided by the Data Science Bowl 2018 (DSB2018) segmentation challenge, consists of $670$ images of nuclei from different modalities (brightfield vs. fluorescence) with an image size of $256\times256$.

\paragraph{Evaluation Criteria}We evaluate the performance of PMR-Net and other methods using three widely-used metrics: \bm$Acc$, \bm$AUC$, and \bm$IoU$. \bm$Acc$ indicates the proportion of correct predictions in the total data:
\begin{equation}\label{8}
Acc= \frac{TP+TN}{TP+TN+FP+FN},
\end{equation}
where \bm$TP$, \bm$TN$, \bm$FP$ and \bm$FN$ denote the number of true positives, true negatives, false positives, and false negatives, respectively.

\bm$AUC$ represents the probability that the predicted positive example will be ranked ahead of the negative example. In the curve, the horizontal axis is denoted as \bm$FPR$, and the vertical axis is represented as \bm$TPR$. \bm$TPR$ and \bm$FPR$ are respectively calculated as follows:
\begin{equation}\label{9}
TPR= \frac{TP}{TP+FN},
\end{equation}
\begin{equation}\label{10}
FPR= \frac{FP}{TN+FP}.
\end{equation}

\bm$IoU$ is the intersection over union, which always measures the degree of overlap between the segmentation result and the ground truth. \bm$IoU$ is calculated in (11):
\begin{equation}\label{11}
IoU= \frac{TP}{TP+FP+FN}.
\end{equation}

\bm$Acc$ and \bm$IoU$ are between $0.0$ and $1.0$, and \bm$AUC$ is between $0.5$ and $1.0$. The closer the experimental result score is to $1.0$, the better the model performance is. Otherwise, the smaller the score is, the worse the model performance is.

\paragraph{Implementation details}All the experiments are performed using an NVIDIA GeForce RTX 2080Ti GPUs with 11 GB memory, Intel(R) Xeon(R) Gold 6226R CPU with 32GB memory, and Ubuntu 16.04.10. PMR-Net is implemented based on the PyTorch repository. The Adam optimizer is used in the PMR-Net training process. The maximum number of training epochs is set to $150$. The batch size is $4$, the initial learning rate is set to $10^{-4}$, the momentum is $0.9$ and the weight decay value is $10^{-5}$.

The lung, retinal vessels (STARE), and cell nuclei datasets were randomly divided into training and test sets according to the ratio of $8:2$. For the skin lesions and retinal vessels (DRIVE) datasets, the division of training and test sets is consistent with the official description. The images on the skin lesions dataset need to be downsampled to $512\times512$ due to memory limitations. For lung, retinal vessels (DRIVE), retinal vessels (STARE) and cell nuclei datasets, the training and test sets are augmented in the same way to prevent over-fitting due to few training samples during model training. The number of samples is augmented to $8$ times of the original dataset. The data augment methods used mainly include: horizontal flipping, vertical flipping, diagonal flipping, HSV color space transformation and image shifting.

\subsection{Comparisons to the State-of-the-Arts}
To evaluate the segmentation performance of PMR-Net, we conduct the experiments on the above datasets and compared with state-of-the-art methods including UNet++ [12], CE-Net [31], DeepLabv3+ [30], MSRF-Net [9], and TransUNet [20]. Especially, CE-Net and Deeplabv3+ employ the pre-trained ResNet (ResNet34 and ResNet101) as the backbone to extract features. In addition, to verify that PMR-Net has good flexibility in structural design, we set the number of network layers of PMR-Net to $4$ and the number of parallel multi-resolution branches to $4$ to obtain a new network with less parameters, which is named PMR-Net-Tiny. In next, we demonstrate the segmentation results of PMR-Net, PMR-Net-Tiny and other state-of-the-art methods on five public available benchmarks.
\subsubsection{Lung segmentation}
Lung segmentation experiments are performed using PMR-Net, CE-Net, UNet++, DeepLabv3+, MSRF-Net, and TransUNet, and the segmentation results are shown in Figure 5. In Figure 5, we can observe that PMR-Net results in more accurate segmentation results of lung edges than other methods. UNet++ and CE-Net generate obvious noise interference in the lung regions due to not fusing the semantic features with the recovered local features during the decoding process. DeepLabv3+ has a slight improvement over UNet++ and CE-Net, but the segmentation effect at the lung edges is still inaccurate. The main reason is that DeepLabv3+ only executes one feature fusion operation during the decoding process. MSRF-Net is only lower than PMR-Net in term of $Acc$ and $IoU$, which is mainly due to the effective fusion of multi-scale features using DSDF module. TransUNet has slightly under-segmentation at the contour of the lung. In the decoding process, PMR-Net continuously fuses high-level context features with the local features recovered by decoding, which makes the segmentation convergence effect at the edge of the lungs better. The quantitative evaluation on the lung dataset are shown in Table $2$. In Table $2$, it can be seen that PMR-Net performs better than the comparative methods on the lung segmentation task. In addition, Although PMR-Net-Tiny has less parameters than the comparative methods, PMR-Net-Tiny can also achieve good segmentation results. Therefore, PMR-Net is an effective and general network architecture for medical image segmentation. 

\begin{figure*}[htbp]
	\centering
	\includegraphics[width=1.0\linewidth]{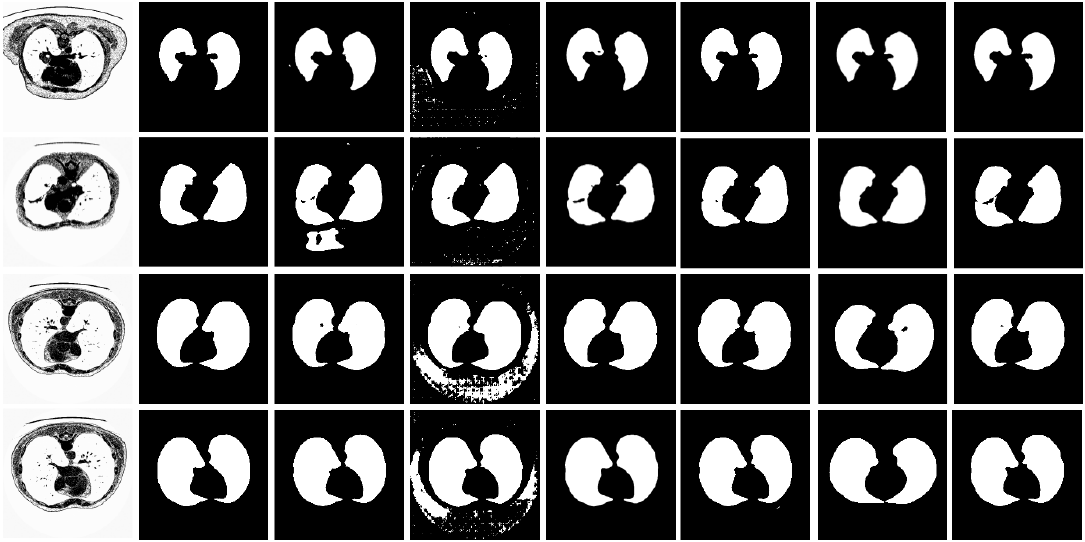}
	\scriptsize{Image~~~~~~Ground Truth~~~UNet++~~~CE-Net~~~DeepLabv3+~~~MSRF-Net~~~TransUNet~~~PMR-Net}
	\caption{The segmentation results of all methods on the lung dataset.}
\label{fig5}
\end{figure*}

\begin{table}[h]\centering
\caption{Performance evaluation of all methods on the lung dataset. The best values are in bold.}
\label{table}
\begin{tabular}{cccc}
\hline
Methods&$Acc$(Mean±Std)&$AUC$(Mean±Std)&$IoU$(Mean±Std)\\
\hline
UNet++ [12]&0.993±0.003&0.999±0.001&0.968±0.014\\
CE-Net [31]&0.983±0.014&0.996±0.003&0.928±0.040\\
DeepLabv3+ [30]&0.994±0.002&0.999±0.000&0.974±0.010\\
MSRF-Net [9]&0.994±0.002&0.990±0.005&0.976±0.011\\
TransUNet [20]&0.993±0.011&0.999±0.000&0.972±0.029\\
PMR-Net&\textbf{0.995±0.002}&\textbf{0.999±0.000}&\textbf{0.978±0.009}\\
PMR-Net-Tiny&0.994±0.003&0.999±0.000&0.972±0.010\\
\hline
\end{tabular}
\label{tab2}
\end{table}

\subsubsection{Retinal vessels segmentation}
For the retinal vessel segmentation task, segmentation experiments are performed on the DRIVE and STARE datasets, respectively. The segmentation results on datasets DRIVE and STARE are visually shown in Figures 6 and 7, respectively. In Figures 6 and 7, CE-Net obtains multiple discontinuous segmentation results at the tail of the retinal vessel, the segmented vessels are not clear, and there are many interference in the segmentation results. DeepLabv3+ has discontinuity at the segmentation results, and the contour of the retinal vessel is generally over-segmented. UNet++ uses long and short skip connections to fuse features, so the segmentation results are better than CE-Net and DeepLabv3+. However, in the process of decoding, UNet++ does not fuse semantic features with local features, resulting in the loss of context features at the edges and discontinuous segmentation at the tails of retinal vessels. Compared with UNet++, PMR-Net continuously supplements high-level global context features in the decoding process. MSRF-Net leads to the loss of small blood vessel objects in the retinal vessel segmentation task. TransUNet has a weak segmentation effect at the edge of the object due to the lack of extraction and fusion of neighborhood feature information. The decoding process of PMR-Net not only restores pixel-level features, but also retains neighborhood features of pixels. Therefore, PMR-Net can retain more complete details in the segmentation results, and obtain more accurate segmentation results.

\begin{figure*}[!t]
\centerline{\includegraphics[width=1\linewidth]{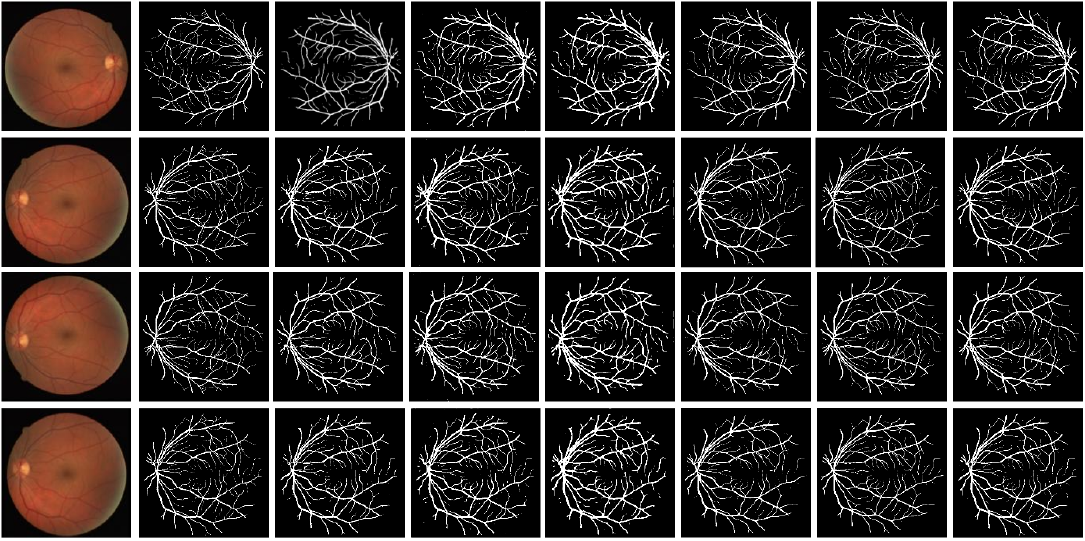}}
\scriptsize{Image~~~~~~Ground Truth~~~UNet++~~~CE-Net~~~DeepLabv3+~~~MSRF-Net~~~TransUNet~~~PMR-Net}
\caption{The segmentation results of all methods on the DRIVE dataset.}
\label{fig6}
\end{figure*}

\begin{table}[h]\centering
\caption{Performance evaluation of all methods on the DRIVE dataset. The best values are in bold.}
\label{table}
\begin{tabular}{cccc}
\hline
Methods&$Acc$(Mean±Std)&$AUC$(Mean±Std)&$IoU$(Mean±Std)\\
\hline
UNet++ [12]&0.968±0.003&0.980±0.006&0.684±0.023\\
CE-Net [31]&0.960±0.003&0.914±0.019&0.646±0.018\\
DeepLabv3+ [30]&0.948±0.005&0.973±0.006&0.589±0.026\\
MSRF-Net [9]&0.966±0.003&0.890±0.022&0.671±0.020\\
TransUNet [20]&0.961±0.005&0.985±0.002&0.668±0.023\\
PMR-Net&\textbf{0.968±0.002}&\textbf{0.986±0.004}&\textbf{0.697±0.020}\\
PMR-Net-Tiny&0.968±0.003&0.964±0.011&0.694±0.020\\
\hline
\end{tabular}
\label{tab3}
\end{table}

\begin{figure*}[!t]
\centerline{\includegraphics[width=1\linewidth]{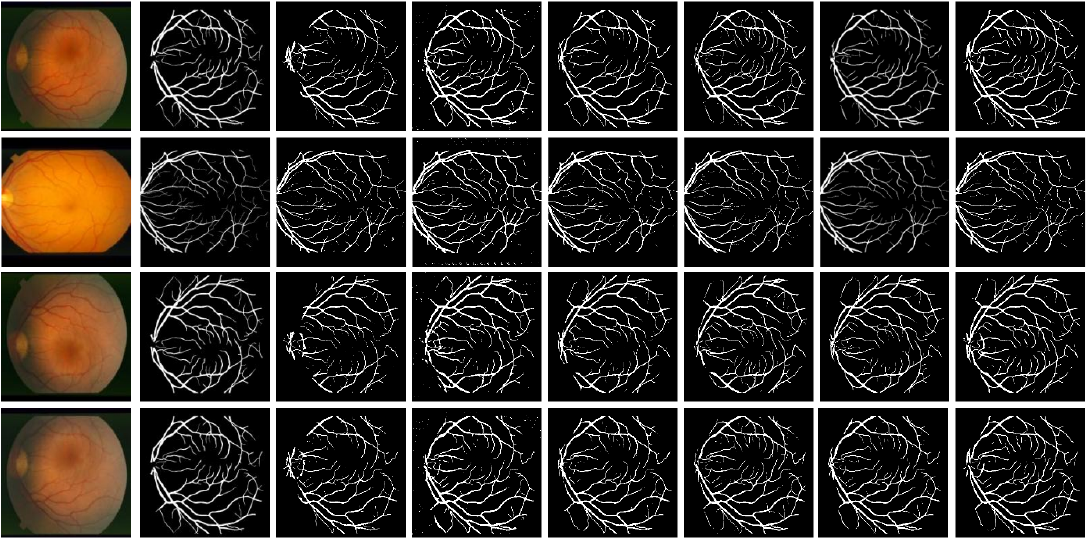}}
\scriptsize{Image~~~~~~Ground Truth~~~UNet++~~~CE-Net~~~DeepLabv3+~~~MSRF-Net~~~TransUNet~~~PMR-Net}
\caption{The segmentation results of all methods on the STARE dataset.}
\label{fig7}
\end{figure*}

\begin{table}[h]\centering
\caption{Performance evaluation of all methods on the STARE dataset. The best values are in bold.}
\label{table}
\begin{tabular}{cccc}
\hline
Methods&$Acc$(Mean±Std)&$AUC$(Mean±Std)&$IoU$(Mean±Std)\\
\hline
UNet++ [12]&0.972±0.010&0.976±0.018&0.661±0.084\\
CE-Net [31]&0.965±0.006&0.922±0.036&0.637±0.040\\
DeepLabv3+ [30]&0.971±0.006&0.983±0.009&0.670±0.047\\
MSRF-Net [9]&0.973±0.007&0.900±0.030&0.683±0.054\\
TransUNet [20]&0.970±0.004&0.988±0.005&0.701±0.056\\
PMR-Net&\textbf{0.975±0.007}&\textbf{0.990±0.005}&\textbf{0.704±0.064}\\
PMR-Net-Tiny&0.972±0.007&0.957±0.026&0.684±0.059\\
\hline
\end{tabular}
\label{tab4}
\end{table}

Table $3$ and Table $4$ show the quantitative evaluation results of the above seven methods on the DRIVE and STARE datasets, respectively. As can be seen from Table $3$, on the DRIVE dataset, PMR-Net scores $0.968$ in the $Acc$, $0.986$ in the $AUC$, and $0.697$ in the $IoU$, ranking first in all three evaluation metrics. In summary, the segmentation performance of PMR-Net is superior to the other five methods on the DRIVE dataset. From Table $4$, we can observe that the three evaluation metrics of PMR-Net, $Acc$, $AUC$ and $IoU$, are better than the other five methods on the STARE dataset. To sum up, PMR-Net achieves better segmentation performance on the retinal vessel segmentation dataset with complex lesion interference.

\subsubsection{Skin lesions segmentation}
The segmentation results of all methods on the lesion skin dataset are shown in Figure 8. In Figure 8, PMR-Net can achieve more accurate segmentation results on the skin lesions dataset compared with the other methods. Especially in the areas with different severity of lesions, the color depth is different, and the lesions with light color are very close to normal skin, resulting in UNet++, CE-Net, and DeepLabv3+ generating more discontinuous segmentation areas, and even more serious incorrect segmentation. Additionally, It can be seen that CE-Net and DeepLabv3+ are better than UNet++. However, since CE-Net and DeepLabv3+ cannot fuse fine-grained local features with global features, the segmentation results are inaccurate in some places where the degree of skin lesions is not obvious. Although MSRF-Net has the good capability of global multi-scale feature representations, MSRF-Net can not still generate the accurate segmentation results on the skin lesions dataset. TransUNet can achieve the good segmentation results due to its powerful global feature representations, but the details at the edge can not be well preserved. PMR-Net can effectively capture global and local features and obtain accurate segmentation results even in regions with different degrees of disease.

\begin{figure*}[!t]
\centerline{\includegraphics[width=1\linewidth]{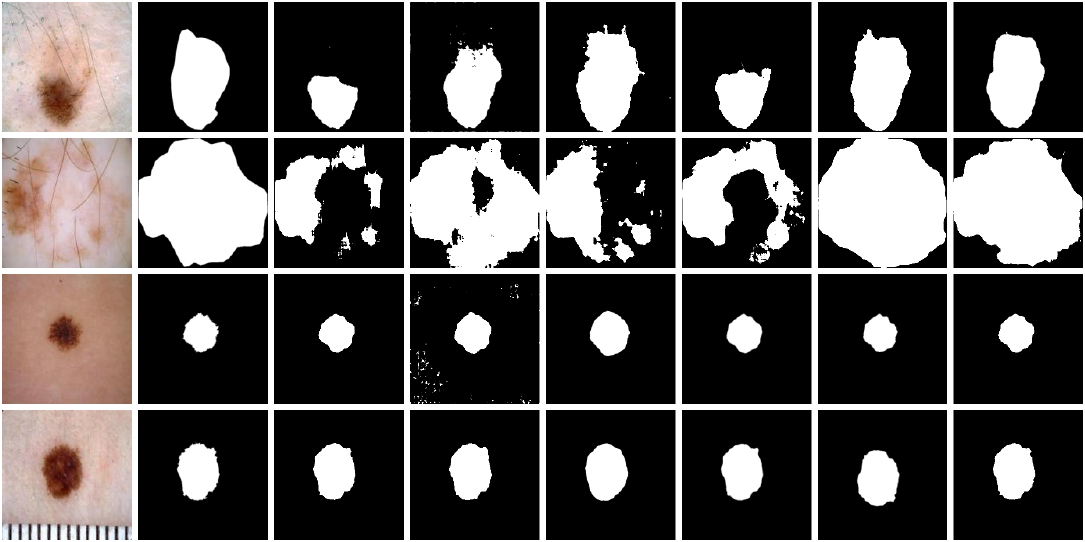}}
\scriptsize{Image~~~~~~Ground Truth~~~UNet++~~~CE-Net~~~DeepLabv3+~~~MSRF-Net~~~TransUNet~~~PMR-Net}
\caption{The segmentation results of all methods on the SL dataset.}
\label{fig8}
\end{figure*}

\begin{table}[h]\centering
\caption{Performance evaluation of all methods on the SL dataset. The best values are in bold.}
\label{table}
\begin{tabular}{cccc}
\hline
Methods&$Acc$(Mean±Std)&$AUC$(Mean±Std)&$IoU$(Mean±Std)\\ 
\hline
UNet++ [12]&0.906±0.107&0.974±0.048&0.726±0.201\\
CE-Net [31]&0.931±0.071&0.960±0.061&0.776±0.137\\
DeepLabv3+ [30]&0.915±0.104&\textbf{0.989±0.025}&0.768±0.152\\
MSRF-Net [9]&0.934±0.082&0.928±0.095&0.798±0.174\\
TransUNet [20]&0.936±0.078&0.988±0.048&0.804±0.137\\
PMR-Net&0.939±0.076&0.988±0.037&0.811±0.138\\
PMR-Net-Tiny&\textbf{0.941±0.077}&0.983±0.044&\textbf{0.819±0.147}\\
\hline
\end{tabular}
\label{tab5}
\end{table}

Furthermore, we quantitatively evaluate the segmentation accuracy and show the evaluation results of seven methods on the skin lesions dataset in Table $5$. In Table $5$, in the lesion skin segmentation task, the $Acc$, $AUC$, and $IoU$ score of PMR-Net is $0.939$, $0.989$, and $0.811$, respectively. Thus, PMR-Net outperforms other state-of-the-art methods on the skin lesions dataset. Interestingly, PMR-Net-Tiny achieves better segmentation results than PMR-Net on the skin lesions dataset, although the parameters of PMR-Net-Tiny is only a quarter of that of PMR-Net. It also reflects that increasing parallel encoder-decoder branches is more important to improve the segmentation performance of the PMR-Net than simply increasing the network depth.

\subsubsection{Cell nuclei segmentation}
Figure 9 visually compares the respective performance on the cell contour dataset, and the segmentation result of PMR-Net is the best. As shown in Table $6$, in the evaluation metrics $Acc$ and $AUC$, PMR-Net scored $0.976$ and $0.993$, respectively, ranking first in these two evaluation metrics. In terms of $IoU$, PMR-Net closely follows UNet++, with a difference of $0.013$. Therefore, PMR-Net can achieve the better segmentation results than these state-of-the-art methods on the DSB2018 dataset.

\begin{figure*}[!t]
\centerline{\includegraphics[width=1\linewidth]{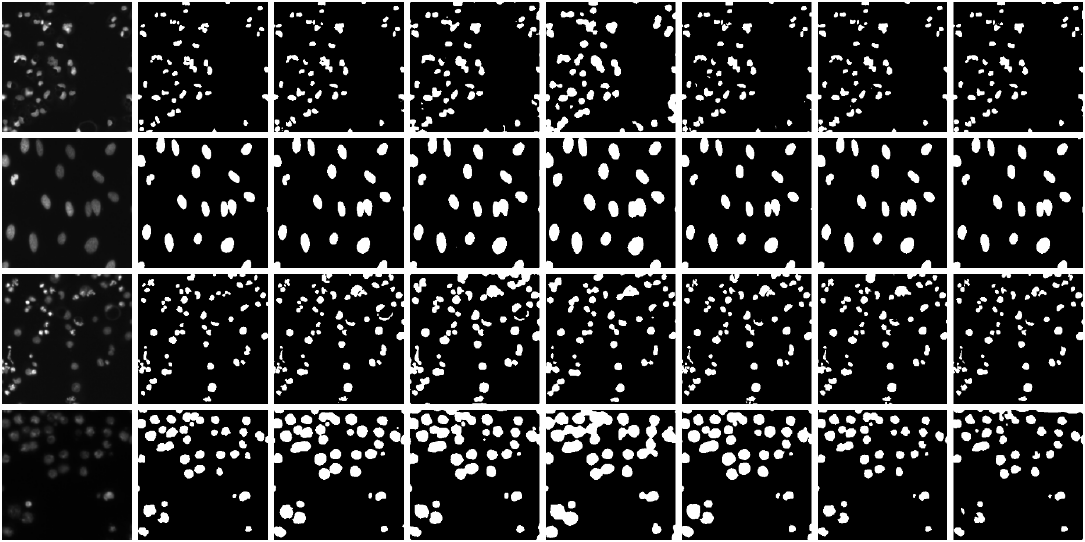}}
\scriptsize{Image~~~~~~Ground Truth~~~UNet++~~~CE-Net~~~DeepLabv3+~~~MSRF-Net~~~TransUNet~~~PMR-Net}
\caption{The segmentation results of all methods on the DSB2018 dataset.}
\label{fig9}
\end{figure*}

\begin{table}[h]\centering
\caption{Performance evaluation of all methods on the DSB2018 dataset. The best values are in bold.}
\label{table}
\begin{tabular}{cccc}
\hline
Methods&$Acc$(Mean±Std)&$AUC$(Mean±Std)&$IoU$(Mean±Std)\\ 
\hline
UNet++ [12]&0.975±0.029&0.991±0.017&\textbf{0.854±0.111}\\
CE-Net [31]&0.960±0.035&0.964±0.081&0.764±0.150\\
DeepLabv3+ [30]&0.950±0.047&0.984±0.030&0.695±0.167\\
MSRF-Net [9]&0.976±0.027&0.952±0.047&0.843±0.122\\
TransUNet [20]&0.976±0.024&0.991±0.015&0.849±0.100\\
PMR-Net&\textbf{0.976±0.029}&\textbf{0.993±0.013}&0.841±0.118\\
PMR-Net-Tiny&0.975±0.029&0.976±0.044&0.834±0.151\\
\hline
\end{tabular}
\label{tab6}
\end{table}

\subsection{Ablation studies}
We investigate the effectiveness of parallel multi-resolution encoder, multi-resolution context encoder and parallel multi-resolution decoder proposed in PMR-Net by carrying out ablation studies on the skin lesion dataset. In the ablation studies, the maximum number of epochs is set to $100$, the learning rate is $10^{-4}$, the momentum is $0.9$, and the weight decay rate is $10^{-5}$.

\begin{table}[h]\centering
\caption{Ablation studies on skin lesion datasets. The best values are in bold.}
\label{table}
\setlength{\tabcolsep}{15pt}
\begin{tabular}{cc}
\hline
Methods&$IoU$(Mean±Std)\\  
\hline
UNet++ (Baseline) [12]&0.726±0.201\\
Baseline + PMR encoder&0.804±0.131\\
Baseline + PMR encoder + PMR decoder&0.805±0.147\\
Baseline + PMR decoder&0.800±0.197\\
PMR-Net (Ours)&\textbf{0.811±0.138}\\
\hline
\end{tabular}
\label{tab7}
\end{table}

\paragraph{Ablation studies for parallel multi-resolution encoder}PMR-Net is inspired by long-short skip connections of UNet++. Thus, UNet++ is chose as the baseline for ablation studies. We replace the encoder of UNet++ with parallel multi-resolution encoder, and the long-short skip connections and decoder of UNet++ remain unchanged. Rows $1$ and $2$ in Table $7$ show the segmentation results. By adding parallel multi-resolution encoder, the $IoU$ score is $0.804$, which is a $10.7\%$ improvement over the baseline. The results demonstrate the effectiveness of parallel multi-resolution encoder in the segmentation task by fusing local features with global context features during encoding stage.

\paragraph{Ablation studies for multi-resolution context encoder}We replace the encoder and decoder of UNet++ with parallel multi-resolution encoder and parallel multi-resolution decoder, respectively. Compared with this network structure, PMR-Net is the same except for the introduction of multi-resolution context encoder. The segmentation results are shown in Rows $3$ and $5$ of Table $7$. The $IoU$ score of PMR-Net is $0.811$, which is $0.75\%$ higher than the model without multi-resolution context encoder. The results indicate that multi-resolution context encoder proposed in this paper can complement semantic features of different sizes objects, and is a simple but effective module for multi-scale context feature fusion.

\paragraph{Ablation studies for parallel multi-resolution decoder}We firstly perform additional two pooling operations on the high-level features output by the encoder of UNet++ and generate two feature maps with different sizes. Then, we make these two feature maps as the other two inputs of the newly replaced parallel multi-resolution decoder. Finally, we replace the decoder of UNet++ with parallel multi-resolution decoder. The segmentation results are shown in Rows $1$ and $4$ of Table $7$. After adding parallel multi-resolution decoder, the $IoU$ is $0.800$, which is $10.2\%$ higher than that of the baseline. The results demonstrate that parallel multi-resolution decoder proposed in this paper can guide the decoding process to effectively recover the local features by using plenty of global context features, and achieve better segmentation results at the edges of objects.

\section{Discussion}
In this paper, we study the network architecture of PMR-Net that composed of different combinations of network depths and parallel encoder-decoder branches. The PMR-Net can not only extract high-level context semantic features in the deep network layers, but also supplement parallel encoder-decoder branches in each layer of the network to extract and integrate multi-scale features, so as to achieve accurate and complete segmentation results for objects with different sizes. In this section, we first discuss the trade-off between the segmentation performance and complexity of the PMR-Net. Then, we investigate the impacts of the number of parallel encoder-decoder branches and network layers on the segmentation performance of the PMR-Net. Finally, we further discuss the flexible scalability of the PMR-Net.

\subsection{The trade-off between segmentation performance and complexity}
There are two key factors that need to be considered in the process of designing the network structure. On the one hand, it is necessary to indicate the effectiveness of supplementing parallel encoder-decoder branches to the baseline model (UNet++). On the other hand, we also consider both the computation and memory usage of the PMR-Net to give a general scheme to adjust the number of parallel encoder-decoder branches. Therefore, we conduct the experiments on the skin lesion dataset. The experimental results are shown in Table $8$. In the first row of Table $8$, the baseline model, UNet++, is with five layers and only one encoding and decoding branch.

\begin{table}[!htb]\centering
\caption{The effect of increasing the parallel encoder-decoder branches on the segmentation performance. The best values are in bold.}
\label{table}
\begin{tabular}{ccccccc}
\hline
\makecell[t]{Lay- \\ ers}&\makecell[t]{Bran- \\ ches}&\makecell[t]{$IoU$ \\ (Mean±Std) }&\makecell[t]{ $AUC$ \\ (Mean±Std)}&\makecell[t]{$Acc$ \\ (Mean±Std)}&\makecell[t]{Param- \\ eters/M}&\makecell[t]{Inference \\ time/ms}\\  
\hline
5&1&0.726±0.201&0.974±0.048&0.906±0.107&\textbf{9.16}&\textbf{33.10}\\
5&2&0.806±0.145&0.981±0.043&0.935±0.083&28.95&43.39\\
5&3&0.811±0.138&\textbf{0.988±0.037}&0.939±0.076&31.20&49.02\\
5&4&\textbf{0.826±0.144}&0.968±0.078&\textbf{0.944±0.079}&33.45&56.40\\
6&3&0.816±0.153&0.987±0.042&0.940±0.078&125.20&72.14\\
\hline
\end{tabular}
\label{tab8}
\end{table}

In Table $8$, the number of parallel encoder-decoder branches of the PMR-Net is set to $2$, the parameters of the PMR-Net is increased by nearly three times and the inference time of the PMR-Net is increased by more than 10ms. The PMR-Net extracts and integrates multi-scale features by increasing the number of branches, which effectively improves the segmentation performance, but also increases the parameters and inference time of the PMR-Net to a certain extent. When the number of parallel encoder-decoder branches is increased continuely, the segmentation performance of the PMR-Net can still be improved. Meanwhile, the parameters and inference time of the PMR-Net increase only slightly. Therefore, for some scenarios with requirement of high segmentation accuracy, we increase parallel encoder-decoder branches and remain the layers of the PMR-Net unchanged, which can effectively improve the segmentation performance and slightly increase the parameters and inference time. Moreover, when the number of parallel encoder-decoder branches of the PMR-Net is increased to $3$ or $4$, the PMR-Net achieves a good balance in the segmentation performance, parameters and inference time.

To verify the impact of deepening the network layers on the segmentation performance, the number of parallel encoder-decoder branches of PMR-Net is fixed to $3$, and the number of layers is increased from $5$ to $6$. Comparing the third and fifth rows in Table $8$, it can be found that the parameters and inference time of the PMR-Net increase significantly due to increasing the number of network layers, but the $IoU$, $AUC$ and $Acc$ can not improve significantly. Therefore, increasing the number of network layers is not an effective solution to improve the segmentation performance of the PMR-Net.

Additionally, the proposed PMR-Net consists of five layers and three parallel encoder-decoder branches, which achieves a good balance between the segmentation performance and the number of parameters and is much better than UNet++ for segmenting objects with different sizes. 
\subsection{Relationship between parallel encoder-decoder branches and network depth}
In this section, the relationship between the number of parallel encoder-decoder branches and the depth of PMR-Net is discussed on the premise of considering the segmentation accuracy, parameters and inference time. We try to increase parallel encoder-decoder branches and appropriately reduce network layers, and carry out segmentation experiments on skin lesion datasets. The experimental results are shown in Table $9$.

\begin{table}[h]\centering
\caption{The effect of the parallel encoder-decoder branches and network depth on the segmentation performance. The best values are in bold.}
\label{table}
\begin{tabular}{ccccccc}
\hline
\makecell[t]{Lay- \\ ers}&\makecell[t]{Bran- \\ ches}&\makecell[t]{$IoU$ \\ (Mean±Std) }&\makecell[t]{ $AUC$ \\ (Mean±Std)}&\makecell[t]{$Acc$ \\ (Mean±Std)}&\makecell[t]{Param- \\ eters/M}&\makecell[t]{Inference \\ time/ms}\\ 
\hline
5&1&0.726±0.201&0.974±0.048&0.906±0.107&9.16&33.10\\
4&2&0.785±0.173&0.973±0.060&0.929±0.095&6.22&33.04\\
4&3&0.803±0.146&0.977±0.052&0.936±0.079&7.71&37.36\\
4&4&\textbf{0.819±0.147}&\textbf{0.983±0.044}&\textbf{0.941±0.077}&8.27&41.29\\
3&3&0.765±0.189&0.973±0.048&0.922±0.101&\textbf{1.85}&\textbf{25.24}\\
\hline
\end{tabular}
\label{tab9}
\end{table}

It can be seen from the first, second, third and fourth rows in Table $9$ that after reducing the number of network layers from $5$ to $4$, $IoU$, $AUC$ and $Acc$ of the PMR-Net have been effectively improved by successively increasing parallel encoder-decoder branches. Meanwhile, the parameters of the PMR-Net are still less than those of the UNet++. Increasing parallel encoder-decoder branches brings less additional computation and slightly increases the inference time.

By comparing the third and fifth rows in Table $9$, it can be found that after reducing the number of network layers from $4$ to $3$, $IoU$, $AUC$ and $Acc$ are decreased, and the parameters and inference time are greatly reduced. The parameters of PMR-Net is reduced from $7.71$M to $1.85$M, and the inference time of PMR-Net is reduced from $37.36$ms to $25.24$ms. Besides, when the number of network layers is $4$ and the number of parallel encoder-decoder branches is increased to $4$, $IoU$, $AUC$ and $Acc$ are increased, and the parameters and inference time are also increased slightly. Therefore, the complexity of PMR-Net can be decreased by reducing network layers and increasing parallel encoder-decoder branches at the cost of decreasing the segmentation accuracy a little. In summary, PMR-Net can also provide an effective solution for medical image segmentation in some resource constrained scenarios.

\subsection{Flexibility of PMR-Net}
The proposed PMR-Net has flexible structure scalability, which can combine network layers and parallel encoder-decoder branches differently. PMR-Net can be widely used in high-level computer vision tasks, such as image segmentation and object detection. For the task of image segmentation, the three important components proposed in PMR-Net, parallel multi-resolution encoder, multi-resolution context encoder and parallel multi-resolution decoder, can not only be directly applied to UNet++, but also be applied to other U-shaped networks in a plug-and-play manner. For the object detection task, the parallel multi-resolution encoder can be used as the backbone network to extract image multi-scale features.

In addition, PMR-Net can also introduce other modules that can enhance multi-scale feature fusion, such as attention mechanism. In the multi-scale feature fusion between parallel multi-resolution encoder branches of PMR-Net, the attention mechanism is introduced to calculate the spatial attention weights between multi-scale features, which can further enhance the feature fusion with the same semantic information and suppress the interference between different semantic information.

\section{Conclusion}
In this paper, we have investigated the potentials of multi-scale feature representation and global context features preservation on medical image segmentation by designing three important modules: parallel multi-resolution encoder, multi-resolution context encoder and parallel multi-resolution decoder. By plugging them into the UNet++ architecture, we have proposed PMR-Net to segment objects with different sizes effectively for medical images. PMR-Net can effectively capture and aggregate local and global features to enhance multi-scale feature representations and improve the accuracy of segmentation results. Extensive experiments demonstrate that PMR-Net is superior to previous state-of-the-art methods on five widely-used medical image segmentation benchmarks. Moreover, PMR-Net is also a general and flexible network framework that can adjust the number of network layers and the number of parallel encoder-decoder branches for different application scenarios.

Generally, different channels of the same scale features have different degrees of importance for the medical image segmentation. Thus, it is important to extract useful features of channel dimension according to the practical requirements of medical image segmentation task. Meanwhile, it should be considered to leverage an effective long-distance dependence between local and global features, which is more conducive to segment the objects with blurred edges. Therefore, we will further study the feature extraction of the same scale and the feature fusion of different scales. In addition, PMR-Net will be applied to 3D medical image segmentation in the future work.

\section*{Acknowledgments}
This work is partly supported by National Natural Science Foundation of China (Nos. 61861024, 61871259, and 61762058), Natural Science Foundation of Gansu Province of China (Nos. 20JR5RA404, and 21JR7RA282), Natural Science Basic Research Program of Shaanxi(Nos. 2021JC-47, 2022JQ-634, and 2022JQ-018), Key Research and Development Program of Shaanxi (Nos. 2022GY-436, and 2021ZDLGY08-07), and Shaanxi Joint Laboratory of Artificial Intelligence (No. 2020SS-03).



\end{document}